\documentclass{INTERSPEECH2023}
\usepackage{amsmath,graphicx,amssymb,delarray,subfigure,verbatim, bbding,boondox-calo}
\usepackage{diagbox, makecell}
\usepackage{bbding}
\usepackage{multirow}
\usepackage{color}
\usepackage{hyperref}

\title{A High Fidelity and Low Complexity Neural Audio Coding}

\name{Wenzhe Liu$^{1}$, Wei Xiao$^{1}$, Meng Wang$^{1}$, Shan Yang$^{2}$, Yupeng Shi$^{1}$, Yuyong Kang$^{1}$, Dan Su$^{2}$, Shidong Shang$^{1}$, Dong Yu$^{2}$}

\address{
	$^1$Tencent Ethereal Audio Lab\\
	$^2$Tencent AI Lab
}
\email{ \{wenzheliu, denniswxiao, markuswang, shaanyang, yupengshi, yuyongkang, dansu, simeonshang, dyu\}@tencent.com}

\begin{document}

\maketitle

\begin{abstract}
Audio coding is an essential module in the real-time communication system. Neural audio codecs can compress audio samples with a low bitrate due to the strong modeling and generative capabilities of deep neural networks. To address the poor high-frequency expression and high computational cost and storage consumption, we proposed an integrated framework that utilizes a neural network to model wide-band components and adopts traditional signal processing to compress high-band components according to psychological hearing knowledge. Inspired by auditory perception theory, a perception-based loss function is designed to improve harmonic modeling. Besides, generative adversarial network (GAN) compression is proposed for the first time for neural audio codecs. Our method is superior to prior advanced neural codecs across subjective and objective metrics and allows real-time inference on desktop and mobile.
\end{abstract}
\noindent\textbf{Index Terms}: neural audio/speech codec, perception-based loss, GAN compression, hybrid codec 
%\vspace{-0.2cm}
\section{Introduction}
%\vspace{-0.2cm}
In recent years, video conferencing has become a new normal to help us connect with others anytime and anywhere. Audio coding is one of the most important technologies in real-time communication, which aims to compress the audio signal with the minimum number of bits while retaining its quality. 

Traditional coders can be grouped into two categories, waveform coders and parametric coders~{\cite{stachurski2000combining}}. They remove the redundancy in the audio signal by using the knowledge of psycho-acoustics and speech signal characteristics. Hybrid coders such as Opus~{\cite{valin2012rfc}} and AMR~{\cite{bessette2002adaptive}} combine waveform coders and parametric coders by designing a careful pipeline. 

Although the traditional coder such as Opus is successfully applied in WebRTC, Zoom and YouTube, the speech quality will degrade at very low bitrates. In order to improve the communication experiment and reduce bandwidth costs, neural audio codecs are proposed with the development of deep learning. Deep learning-based audio coders can be classified into two main streams, hybrid neural audio codecs~{\cite{valin2019lpcnet,kleijn2021generative}} and end-to-end audio coding~{\cite{zeghidour2021soundstream, defossez2022high, jiang2022predictive}}. The former leverages generative models as the decoder to generate the speech waveform based on acoustic parameters which are hand-crafted or obtained from conventional parametric coders. For example, LPCNet~{\cite{valin2019lpcnet}} is proposed to improve the speech quality from low-bitrate Opus bitstream thanks to the high-quality speech synthesis capability of generative models. Lyra~{\cite{kleijn2021generative}} feeds quantized log mel spectra to WaveGRU and outperforms Opus at low rates. Another follows the ``encoder-quantizer-decoder'' architecture, where the encoder compresses the speech signal and extracts the latent representation, and the decoder recovers the signal from the quantized features. Compared to hybrid coders, end-to-end coders assign the transformation from acoustic features to latent representations to the encoding side, which reduce the computational cost of the device, and demonstrates better performance. The most popular architecture is VQ-VAE~{\cite{van2017neural}}. SoundStream~{\cite{zeghidour2021soundstream}}/lyra-v2\footnote{https://github.com/google/lyra} and Encodec~{\cite{defossez2022high}} take the time domain samples as input and achieve excellent audio quality with end-to-end adversarial training. Besides, TF-Net~{\cite{jiang2022predictive}} has also made good performance by compressing and modeling in the frequency domain. However, vector quantizer (VQ) raises the storage costs of applications while improving encoding efficiency. 

Among all these approaches, there are two problems: first, the reconstruction speech quality and bandwidth (usually from 8 kHz to 12 kHz) are not satisfactory, and quantized noise and/or spectral artifacts are produced due to limitations of generative models' capabilities. Another problem is that existing neural audio coding frameworks that require superior speech representation usually have high computational and memory costs, and do not meet the real-time requirement.   
To address the above problems, we design a novel audio coding framework that combines deep neural networks and signal processing technologies referring to classical coders. It is well known that high-frequency bands need only a limited number of bits to reconstruct with bandwidth extension (BWE). Therefore, the codec for wideband which consumes more bits is replaced by the neural network. Aiming to overcome the limited spectrum modeling capability of generative models, a perception-based magnitude loss function and postfilter are proposed to constant critical-band energies and suppress the quantization noise. In order to boost the inference speed, lightweight operators are utilized. Meanwhile, GAN compression especially for audio coding is employed for the first time to improve the performance of the lightweight generator. 

In summary, this paper makes the following contributions:
\begin{itemize}
    \item We propose a novel super wideband (SWB) audio codec framework that combines the generative model and classical codecs to achieve superior audio quality but low complexity.
    \item We introduce GAN compression especially for audio coding for the first time, and a series of low-resource operations (such as pooling, repeat, grouped convolution, and the scaler quantizer) replace original modules to reduce compute consumption and storage footprint. 
    \item We design a perception-based loss function and postfilter to further improve audio quality.
    % \item The proposed system has a significant coding efficiency and real-time inference speed, and outperforms the advanced baselines across the subjective and objective evaluation.
\end{itemize}
 
The rest of the paper is organized as follows. In Section~{\ref{system-overview}}, we illustrate the overall diagram. We present the experimental setup in Section~{\ref{experiments}}. The results and analysis are given in Section~{\ref{results-and-analysis}}, and some conclusions are drawn in Section~{\ref{conclusion}}.

\begin{figure*}[t]
	\centering
	\centerline{\includegraphics[width=1.8\columnwidth]{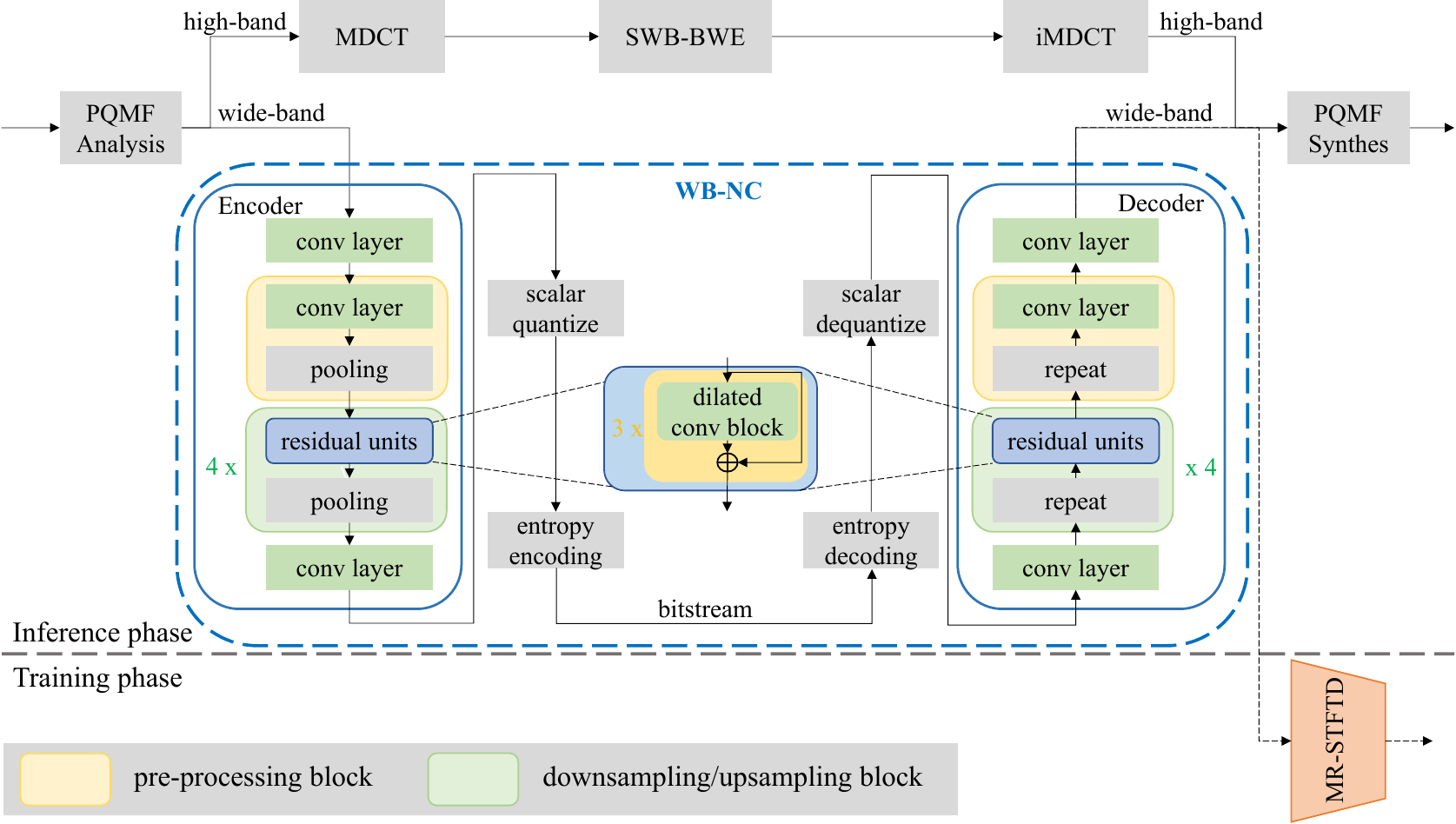}}
	\caption{Overall diagram of the proposed framework.}
	\label{fig:architecture}
	\vspace{-0.6cm}
\end{figure*}

\section{Proposed system}
\label{system-overview}

\subsection{Overview}
\label{overview}
%\vspace{-0.2cm}
The diagram of the proposed codec Penguins is shown in Figure~{\ref{fig:architecture}}. Different from current neural codecs~{\cite{zeghidour2021soundstream, defossez2022high, jiang2022predictive}} that rely completely on deep learning for audio compression, our proposed system consists of two main parts, namely the wideband neural coding (WB-NC) module, and the super wideband bandwidth extension (SWB-BWE) module. It is known that the neural vocoder often causes spectral artifacts, especially at high frequencies due to its weak high-frequency representation capability. On the other hand, conventional coders allocate only a few bitrates to high-frequency bands. Therefore, WB-NC leverages GAN~{\cite{creswell2018generative}} with the advantage of fast inference speed and lightweight networks to model broadband components. SWB-BWE employs a bitrate of 2kbps for the super wideband components encoding and the wideband improvement. The decomposition and synthesis of the wideband signal and the high-band signal are achieved by two-channel quadrature mirror filters (QMF)~{\cite{galand1983design}}. 

The SWB-BWE module refers to \cite{geiser2007bandwidth,wolters2003closer} to transmit the spectrum information in the modified discrete cosine transform (MDCT) domain.
The WB-NC module is composed of a sequence of three
components, as illustrated in Figure~{\ref{fig:architecture}}. First, an encoder is utilized to extract the raw audio waveform to a compressed acoustic representation. After that, a quantizer discretizes the compressed feature with the target bitrate. Finally, a decoder reconstructs the signal from the discrete embeddings. This model is trained in an end-to-end fashion together with discriminators. To alleviate the lack of harmonic expression capability of GANs, a postfilter~{\cite{wong1994coding}} is applied during inference. Each module will be described in detail in the following.

\subsection{Generator}
\label{generator}
The encoder, the quantizer, the decoder, and the post-filter comprise the generator $G_S$ of the WB-NC module. The encoder components consist of a 1D convolutional layer, a pre-processing block, followed by $N$ downsampling blocks. Finally, a 1D convolution with $D$ channels and a kernel size of 3 is used to generate the compressed acoustic representation. The first causal convolutional layer takes 320 samples (20 ms) as the input and transforms them into learnable embeddings. The channel of the first convolution is set to 16. The pre-processing block and downsampling blocks are used to continuously compress the time dimension until it is 1, where the pre-processing block is composed of a causal convolution with 16 channels and a downsampling layer with a downsample rate of 2, and the downsampling block includes three grouped residual units with dilation rate $d = \{1, 3, 9\}$ and a downsampling layer with downsampling rate $r = \{2, 4, 4, 5\}$. The kernel sizes of all convolutional layers are 3, and the number of channels is doubled in the downsampling blocks. Different from previous codecs, average pooling replaces the stride convolution as the downsampling layer due to its extremely low computational cost. Therefore, one acoustic embedding is mapped from $2 \times 2 \times 4 \times 4 \times 5 = 320$ audio samples. 

Between the encoder and the decoder, the quantizer is inserted to turn the output of the encoder into discrete values. Due to large codebooks brought by the residual vector quantization (RVQ) of the neural codecs increasing the storage occupation of RTC software, the scalar quantizer (SQ) is used as an alternative solution to quantize each of the embeddings after 
% the $5 \times tanh(\cdot)$ activate function 
$tanh(\cdot)$
individually without increasing storage requirement. 
% We quantize each embedding which from -5 to 5 with step of 1, and 20 ms audio samples can be represented using $D\cdot \log(11)$ bits. Moreover, entropy coding is used during the inference phase with a range based arithmetic coder, which is lossless compression. 

Following the quantizer, the decoder reconstructs the input speech signal utilizing discrete features, which is similar to the encoder but is the mirror version of the encoder. The decoder component uses upsampling layers instead of downsampling layers. Considering the calculation burden, we replace the transpose convolution commonly used in previous work with the repeat operation, and the upsampling rates are in the reverse order of the downsampling rates. 

% In the inference phase, a post-filter is applied to mitigate the insufficient harmonic representation of the proposed neural codec. More detail information can be found in~{\cite{wong1994coding}}.

\subsection{Discriminator}
\label{discriminator}
Our adversarial training framework relies on multi-resolution STFT-based (MR-STFT) patch discriminators $D_S$, which capture spectral structures of different frequency resolutions. We also investigate other discriminators commonly found in the vocoder such as multi-scale discriminator (MSD)~{\cite{kumar2019melgan}} and multi-period discriminator (MPD)~{\cite{kong2020hifi}}, which have no performance benefits and long training time. 6 different scales are used with FFT points of $K=\{60, 120, 240, 480, 960, 1920\}$. Each discriminator takes the magnitude spectrum and its logarithmic spectrum is concatenated as input and is composed of 7 2D convolution layers with a kernel size of $(3, 3)$ and a stride of $(1, 1)$ or $(2, 2)$. 
% Weight normalization and LeakyReLU are applied sequentially after each convolution layer except the last one.

\subsection{Training loss functions}
\label{loss}
The training loss is a combination of a reconstruction loss term, an adversarial loss term, and the feature match loss term.
where the reconstruction loss consists of a multi-resolution full-band and subband short-time Fourier transform (STFT) loss and a proposed perception-based magnitude (PM) loss. 

For the multi-resolution STFT loss, we minimize the spectral convergence loss~{\cite{arik2018fast}} and the L1 distance in the logarithmic magnitude spectral domain with different FFT analysis parameters, which can be written as:
\begin{equation}
\label{eq:mrstftloss}
\mathcal{L}_{s}(X) = \sum_{r}{\left(||log(X_r)-log(\widehat{X}_r)||_1 + \frac{||X_r-\widehat{X}_r||_F}{||\widehat{X}_r||_F}\right)},
\end{equation}
where $X_r$ and $\widehat{X}_r$ are the spectrum of the clean speech and the predict waveform with FFT-point of $2^r$. 

For full-band and subband cases, their corresponding loss functions are defined as $\mathcal{L}_{s}(S)$ and $\mathcal{L}_{s}(S^{sub})$, where $S$ and $S^{sub}$ represent the magnitude spectrum of the complete signal $s$ and subband signals $s^{sub}$ decomposed by pseudo-quadrature mirror filters (PQMF), respectively. 
It is known that always constant energy in the critical bands is important for the codec, and perceptible quantization noise is often due to the noise energy in spectral valleys being in excess of the auditory masking. The PM loss function is proposed to solve these problems. For the former, $\mathcal{L}_{s}(E)$ focuses on the energy in the designed auditory-perceptual scale, where $E$ is the equivalent rectangular bandwidth (ERB) spectra. For the latter, $\mathcal{L}_{val}=\sum_{r=9}^{11}{M\cdot S_r^p \cdot ||(S_r-\widehat{S}_r)||_1}$ pays more attention to spectral valleys, where $S_r^p$ is a weight implemented by $p$-law compression of the spectrum, which can enlarge small values and reduce spectral peaks when $p
\textless 0$. $M$ is a mask of voice activate detection (VAD) to avoid scaling up the values in silence frames. The loss function can be expressed as:
\begin{equation}
\label{eq:perceploss}
\mathcal{L}_{pm} = \mathcal{L}_{spec}(E) + \mathcal{L}_{valley}.
\end{equation}

The adversarial loss follows LS-GAN~{\cite{mao2017least}}, which forces the generator to fool the discriminator when training the generator. On the other hand, when the discriminator is trained, the LS-GAN helps the discriminator recognize the clean samples as 1, and the samples coded by WB-NC as 0. The loss functions for the generator $G_S$ and the discriminator $D_S$ are expressed as:
\begin{gather}
\label{adv}
 \mathcal{L}_{adv} = \mathbb{E}\left[(1-D_S(\hat{s}))^2\right],\\
\mathcal{L}_{D_S} = \mathbb{E}\left[(D_S(s)-1)^2 +(D_S(\hat{s}))^2\right], 
\end{gather}

 Additionally, the feature match loss~{\cite{kumar2019melgan}} is calculated to minimize the L1 distance between feature maps of the discriminator for real and generated audio, which has been demonstrated to be effective in previous work.
 \begin{equation}
\label{eq:feat}
\mathcal{L}_{f} = \mathbb{E}\left[ \frac{1}{L}\sum_{l=0}^{L-1}{|D_S^{l}(s) - D_S^{l}(\hat{s})|} \right],
\end{equation}
which $L$ denotes the number of the discriminator's layer.

The overall generator loss is a weighted sum of the above loss terms:
 \begin{equation}
\label{eq:feat}
\mathcal{L}_{G_S} = \mathcal{L}_{s}(S) + \mathcal{L}_{s}(S^{sub}) + \lambda_{pe}\cdot\mathcal{L}_{pm} + \lambda_{adv}\cdot\mathcal{L}_{adv} + \lambda_{f}\cdot\mathcal{L}_{f}.
\end{equation}
\subsection{GAN compression}
\label{gan compression}
As can be seen from the previous section, the computational complexity and number of parameters of our proposed model are far less than those of previous work. A series of low-computational-cost alternative operations are applied, which inevitably degrade the performance of the speech reconstruction. In addition to the improvement of speech coding quality using well-designed loss functions mentioned in section~{\ref{loss}}, we also employ model compression methods namely \textit{GAN compression}~{\cite{li2020gan,liu2021content}} to improve the performance of the proposed model. The adversarial training of GAN is unstable since its generator and discriminator are updated alternately as opponents compete. Consequently, GAN requires a specific compression method, in difference to CNN compression. To the best of our knowledge, this is the first proposed GAN-knowledge distillation (KD) algorithm specific to audio codecs. We pre-train a high-computational complexity model (teacher model $\{G_T, D_T\}$) as the teacher, which transfers the information to the proposed model (student model $\{G_S, D_S\}$) with GAN-KD. In the KD stage, the parameters of the teacher model's discriminator are used to initialize the discriminator of the lightweight model. We calculate the distance between the intermediate outputs of $D_T$ and that of $D_S$, and the features of these two models' generators are also calculated in the distillation loss. The KD loss can be expressed as:
\begin{equation}
\label{eqn6}
\mathcal{L}_{KD} = \lambda_{G}\sum_i{||O_i^{G_T}-O_i^{G_S}||_1} + \lambda_{D}\sum_i{||O_i^{D_T}-O_i^{D_S}||_1}
\end{equation}
where $O_i^{\Theta}$ is the output of $i$-th layer of the model $\Theta\in\{G_S;D_S;G_T,D_T\}$.
Additionally, we design a training schedule for GAN-KD for stable and effective training. The learning rate for the generator is updated with a cosine decay from 2e-4, and that for the discriminator is always 1e-5. The weight decay is scheduled with an increasing cosine schedule
leading to a larger regularization for the later steps.

\section{Experiments}
\label{experiments}
% \vspace{-0.2cm}
\subsection{Dataset}
\label{dataset}
% \vspace{-0.2cm}
In the experiments, we use the LibriTTS (English)~{\cite{zen2019libritts}}, DNS Challenge (English and Mandarin) ~{\cite{reddy2020interspeech}, and private datasets (English and Mandarin) as the speech corpus. The noise clips from the DNS challenge are utilized as the noise set. MIR-1k~{\cite{hsu2009improvement}} and FMA~{\cite{defferrard2016fma}} datasets are used as the music set. All clips are resampled at 16 kHz, and the packet loss case is simulated for packet loss concealment (PLC). 

The evaluation metrics include objective and subjective evaluations. For the objective evaluation, the test set is from ITU-T P.501 Annex C\footnote{https://handle.itu.int/11.1002/1000/14271}, which is a multilingual dataset prepared for ITU-T P.800 conformant applications and perceptual-based objective speech quality prediction. Please note that we only trained the model with English and Mandarin utterances, but evaluated under 8 languages. 
ViSQOL~{\cite{hines2012visqol}} is chosen as the objective evaluation metric because it has been used in previous literature~{\cite{zeghidour2021soundstream, defossez2022high}}, although they can only assess wideband speech. Besides, the more standard and commonly used POLQA~{\cite{beerends2013perceptual}} has also been added to evaluate speech quality. Moreover, eSTOI~{\cite{jensen2016algorithm}} is used as the metric of intelligibility. 
Subjective performance is evaluated by an ITU-T P.808 recommended crowdsourced subjective listening test, which is divided into WB and SWB sections. 24 native listeners in Mandarin as subjects for tests, with 12 Mandarin utterances.

\subsection{Training configurations}
\label{training-configurations}
% We train the models for 2000000 steps with the AdamW optimizer. and the ExponentialLR scheduler. 
The batch size is set to 16 with 8 A100 GPUs
% , and each clip is randomly selected for 2 seconds for training
. The FFT points $2^r$ of the MR-STFT are from 512 to 2048. The weights of loss terms $\{\lambda_{pe}, \lambda_{adv}, \lambda_{f}, \lambda_{G}, \lambda_{D}\}$ are $\{2, 1, 20, 30, 10\}$.

\subsection{Baselines}
\label{baselines}
Opus is one of the most popular classical codecs
% Opus is a versatile audio codec 
which ranges from 6 kbps narrowband monophonic audio to 510 kbps full-band stereophonic audio. 
% Opus has been standardized by the IETF and widespread to lots of applications such as Zoom and Tencent Meet. 
We use Opus as the traditional codec baseline.
Lyra-v2 is based on a famous end-to-end audio codec namely SoundStream, which supports bitrates at 3.2 kbps and 6 kbps for 32 kHz audio samples. Like Lyra-v2, Encodec is also a state-of-the-art neural audio coder which compresses 24 kHz audio samples to 1.5-24 kbps. These models are chosen as the deep learning-based baselines and implemented officially.

\section{Results and analysis}
\label{results-and-analysis}

\renewcommand\arraystretch{0.84}
\begin{table}[t]
	\caption{Objective results of different audio codecs. $^\dag$ denotes that Encodec operates audio sampels in 24 kHz sampling rate.}
	\centering
	\large
	\resizebox{0.48\textwidth}{!}{
		\begin{tabular}{l|ccccc}
			\toprule
			Method &Bitrate &Bandwidth &POLQA &ViSQOL &eSTOI(\%) \\
                \hline
                Opus &6 kbps &WB &1.82 &2.03 &58.45\\
                Opus &8 kbps &WB &2.96 &3.46 &76.03\\
                Opus &10 kbps &SWB &3.26 &3.71 &77.93\\
                Opus &16 kbps &WB &4.26 &4.30 &85.30\\
                \hline
                Lyra-v2 &9.2 kbps &WB &3.49 &3.92 &90.81\\
                Encodec &12 kbps &SWB$^\dag$ &3.72 &4.16 &94.21\\
			\hline
			WB-NC &6 kbps &WB  &3.66 &3.88 &91.59\\
			\quad+PM loss &6 kbps &WB &3.86 &4.19 &93.00\\
			\qquad+GAN-KD &6 kbps &WB &4.06 &4.22 &94.63\\
                \quad\qquad+SWB-BWE &8 kbps &SWB &4.16 &4.20 &94.37\\
			\bottomrule
	\end{tabular}
        }
	\label{tbl:ablation}
	\vspace*{-0.39cm}
\end{table}
\renewcommand\arraystretch{0.8}
\begin{table}[t]
	\caption{Real time factor (RTF) for neural audio codecs.}
	\centering
	\tiny %large
	\resizebox{0.32\textwidth}{!}{
		\begin{tabular}{c|c|cc}
			\toprule
                \multirow{2}{*}{Model} &\multirow{2}{*}{Bitrate} &\multicolumn{2}{c}{RTF}\\
                \cline{3-4}
			 & &Enc. &Dec.\\
			\hline
			Lyra-v2 &9.2 kbps &0.015 &0.034 \\
			  Encodec &12 kbps &0.103 &0.094\\
			Proposed &6 kbps &0.032 &0.028 \\
			\bottomrule
	\end{tabular}}
	\label{tbl:rtf}
	\vspace*{-0.39cm}
\end{table}

\subsection{Ablation study}
\label{ablation-study}
Table~{\ref{tbl:ablation}} shows the performance gains obtained by the PM loss, the GAN compression, and the SWB-BWE. One can find that, when the PM loss is applied, it yields 0.20, 0.31 and 1.41\% improvements in POLQA, ViSQOL and eSTOI, respectively. This is because the PM loss emphasizes band energy consistency and spectrum valley noise minimization. 0.20 and 0.10 POLQA gains are achieved by using GAN-KD and SWB-BWE, showing the effect of these strategies.

\subsection{Comparison with other codecs}
\label{comparison-with-advanced-baselines}
The result comparison in terms of POLQA, ViSQOL and eSTOI among different audio codecs is shown in Table~{\ref{tbl:ablation}}. From the results, Several observations can be made. Firstly, neural audio codecs can reduce the bitrates while achieving better performance compared with classical signal processing codecs. Secondly, The proposed system significantly outperforms baselines at a lower bit rate. From Lyra-v2 to our approach, around 0.67, 0.30 and 3.82\% improvements are achieved in terms of POLQA, ViSQOL and eSTOI, indicating the
superiority of the proposed framework. Compared with Encodec, our method achieves similar ViSQOL and eSTOI, but 0.44 POLQA gain. 

\vspace{-0.2cm}
\begin{figure}[t]
	\centering
	\centerline{\includegraphics[width=0.95\columnwidth]{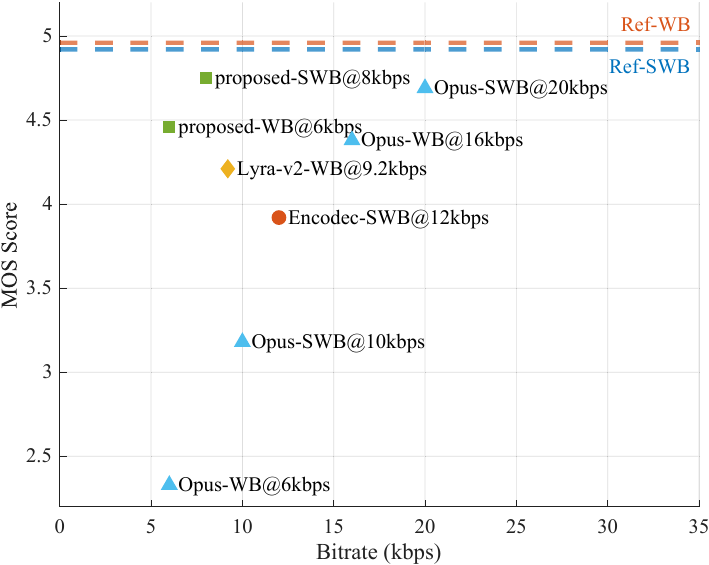}}
	\caption{Subjective evaluation results. The red and blue dotted lines denote the scores of the reference in the WB case and SWB case, respectively.}
	\label{fig:dnsmos}
	\vspace{-0.4cm}
\end{figure}
\subsection{Subjective evaluation}
\label{subject-eval}
The subjective evaluation results of different codecs are reported in Figure~{\ref{fig:dnsmos}} , where *-WB and *-SWB represent wideband and super wideband subjective listening testings. We find that the proposed system outperforms advanced neural audio coders, which is consistent with objective evaluation. Note that the performance of the proposed framework is close to Opus at 20 kbps which is the recommended bit rate for wideband speech communication, demonstrating the excellent performance of our approach. 

\subsection{Inference speed}
\label{inference speed}
% \vspace{-0.2cm}
In Table~{\ref{tbl:rtf}}, we present the real-time factors (RTF) of neural audio codecs on a single thread of a MacBook Pro 2019 (2.6 GHZ i7). Compared with Encodec, the proposed model is better in terms of RTF. Meanwhile, the inference speed of the proposed model is comparable to that of Lyra-v2. We intended the decoder to have a lower computational complexity since it is common to send only one bitstream but possibly decode multiple bitstreams during conferencing. Besides, the inference speed of the proposed method optimized by NCNN\footnote{https://github.com/Tencent/ncnn} on iPhone XS is evaluated, The RTFs of the encoder and the decoder are 0.017 and 0.012, showing the proposed framework is also sufficient to run in real-time on mobile.
% alone or even in a conferencing system such as WebRTC.

\section{Conclusions}
\label{conclusion}
% \vspace{-0.1cm}
In this paper, 
we propose 
% a lightweight audio codec for efficient compression and high-fidelity reconstruction. 
% We adopt 
a hybrid codec that combines the traditional codec and the neural codec to implement efficient compression and high-fidelity reconstruction. Low computational cost operators, perception-based loss, and GAN compression are proposed to improve the generation capability of the GAN. Experimental results show that our framework outperforms prior arts across the objective
and subjective evaluation. 

% \section{Acknowledgement}
% \label{acknowledgement}

\vfill\pagebreak
\bibliographystyle{IEEEtran}
\bibliography{refs}

\end{document}